\documentclass[aps,prd,twocolumn,superscriptaddress]{revtex4}
\usepackage{amsmath,epsfig,mathrsfs,graphicx,amssymb,slashed}

\def\be#1\ee{\begin{equation}#1\end{equation}}

\def\rmd{\mathrm d}
\def\rmi{\mathrm i}
\def\rme{\mathrm e}
\newcommand{\ba}{\begin{eqnarray} }
\newcommand{\ea}{\end{eqnarray} }

\begin{document}
\title{Objective realism and freedom of choice in relativistic quantum field theory}
\author{Adam Bednorz}
\email{Adam.Bednorz@fuw.edu.pl}
\affiliation{Faculty of Physics, University of Warsaw
ul. Pasteura 5, PL02-093 Warsaw, Poland}

\date{\today}

\begin{abstract}
Traditional Bell's argument shows that freedom of choice is inconsistent with quantum realism if lack of signaling and sufficiently fast choices and readouts are assumed. While no-signaling alone is a consequence of special relativity, this is not the case of spacetime location of choice and readout. Here we attempt to incorporate freedom of choice into quantum objective realism relying solely on relativistic quantum field theory. We conclude that this is impossible without breaking relativistic invariance and put forward the possibility of signaling faster than light, which cannot be excluded if an ultimate theory violates relativity.
\end{abstract}
\maketitle
\section{INTRODUCTION}

\begin{figure}
\includegraphics[scale=.4]{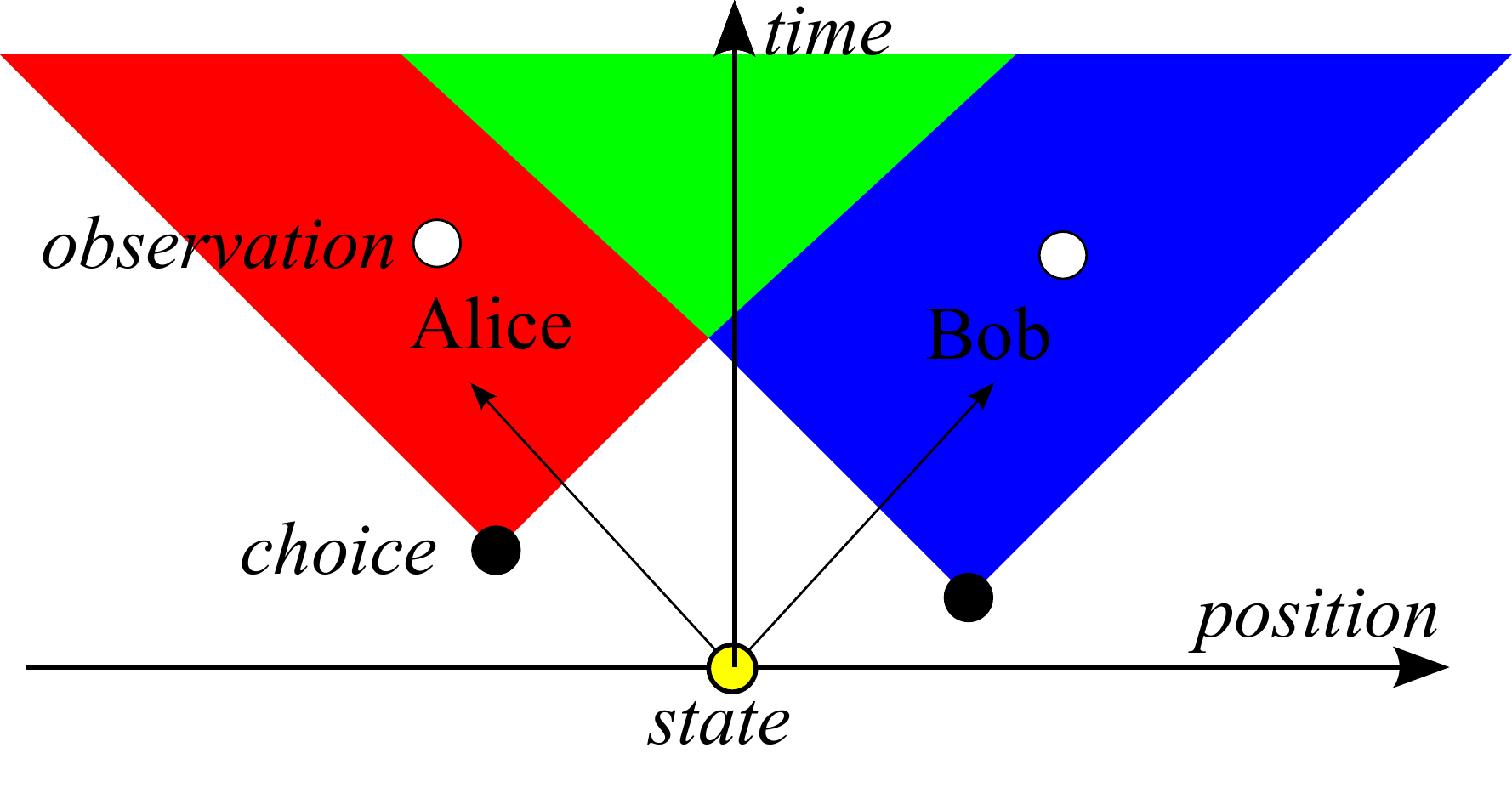}
\caption{Spacetime picture of Bell's assumption. If the two parties, here Alice and Bob, share an entangled state the observation must be completed, before the reach of the signal about the other party's choice (color cones bound by signaling speed -- light in special relativity).}
\label{be1}
\end{figure}

Objective realism means that all physical quantities (e.g. field and currents) have well-defined values at all times and positions, although they may be random. The values are independent of the fact of being observed. Objective realism in the macroscopic world is obvious, but in the microworld it is at best ambiguous due to conceptual problems of the quantum description. 
Moreover, practical and useful physics relies on free choice -- an ability to affect the system in real time.
Freedom of choice means that we are not mere spectators of the world's evolution but can actively change its fate.
Free choice localized in time and space is important in the interpretation of tests of local realism \cite{bell,chsh,eber}.
Incorporating free choice into theory is done by adding some variable parameters (usually localized), meaning a variety of choices.
However, observations for different choices are not always compatible in quantum realism, as shown by Bell theorem (for a particular state and choices) \cite{bell}. The Bell's argument relies on several important assumptions, depicted in Fig. \ref{be1}:
\renewcommand{\labelenumi}{(\roman{enumi})}
\begin{enumerate}
\item Entanglement: existence and stability of a special, nonlocal entangled state, that can be observed by two (or more) separate parties
\item No-signaling: observations are freely chosen and are completed (become sufficiently sharp, with negligible error) before a signal about the other party's choice reaches the observation point
\end{enumerate}
Bell's conclusion is that it is impossible to find a common probability distribution (equivalent to quantum realism) of all outcomes depending only on those choices that can be signaled to them. Both assumptions cannot be directly derived from fully relativistic quantum field theory because the Bell argument works in simplified Hilbert space and reduces to a few basis states. No-signaling could indeed follow from at least axiomatic quantum field theory  \cite{wight} but the point of choice and readout is arbitrary in general. One can easily invalidate the Bell's conclusion by delaying actual observation (or its sharpening) until signals reach its point. Bell theorem has been recently confirmed experimentally \cite{hanson,zein,saew} but of course for no-signaling one assumes special relativity combined with the trust in the times of choices and readouts.

Here  we try to assign joint objective realism for all choices by asking if a common joint positive probability exists and basing it directly on relativistic quantum field theory \cite{peskin}, not Bell's assumptions (so we e.g. do not need to trust the time of choice and readout). We will show that indeed objective realism with free choice cannot stand with both relativistic invariance and quantum theory. It will turn out that it is possible but violating relativistic invariance.
If relativity is to drop, then  binding the assumption of the Bell theorem about compatibility with relativistic no-signaling may be false and there might be signaling faster than light. We show that trying to preserve the speed of light as the signaling speed in a relativity violating theory is misleading if one tries to do it perturbatively. The relativistic signaling limit is simply a nonperturbative property of quantum field theory, and may get falsified in future experiments.

The paper is organized as follows. We start with the general construction of quantum mechanics and field theory with free choice. Next, we state the problem of realism and attempts of quantum construction, insisting on agreement with relativity. Finally, we show that relativistic invariance must be broken, by a perturbative example, and discuss possible consequences, including superluminal signaling. We close the paper with conclusions.

\section{QUANTUM FREEDOM OF CHOICE}

A general construction of quantum observations, satisfying the principle of objective realism, will be completed if the observations  depend  on free-to-choose options, readouts for all options simultaneously are represented by a positive probability.
All events, free choices and measurements will be referred to by time position $x=(x^0=t,\vec{x}=(x^1,x^2,x^3))$
(time $t$, spatial coordinates $\vec{x}$). Speed of light $c$ and Planck constant $\hbar$ are $1$ in our units.
Given the initial state of the system (the Universe) and it dynamics, Hermitian Hamiltonian $\hat{H}$,
the free choice $a$ means a parametric decision to modify the dynamics by an extra term in the Hamiltonian $\hat{H}_a(x)$. If this term is nonzero only around a specific point in spacetime, then we can claim it as localized which is important e.g. in the Bell theorem. However, for our considerations $\hat{H}_a(x)$ will be completely general. There can be many such defined choices, $a,b,c,...$ 
We denote $\hat{O}(x)$ an observable (Hermitian) in the Heisenberg picture with respect to the original Hamiltonian, while for $\hat{O}_a(x)$, $\hat{O}_b(x)$, $\hat{O}_{ab}(x)$ we add the choice-dependent Hamiltonian $\hat{H}_a$, $\hat{H}_b$, or $\hat{H}_a+\hat{H}_b$, respectively. 

We can assign $a=0$ for the null passive choice, $\hat{H}_a=0$, 
meaning only an internal system's dynamics without changes due to active choices, $a\neq 0$. In field theory, it is convenient to define 
an auxiliary field, e.g. $a(x)$ controlling free choice. The choice is realized by adding to the Hamiltonian $\hat{H}_a=\int \rmd^3 x a(x^0,\vec{x})\hat{V}(x)$,
where $\hat{V}$ is some local operator. Quantum field theory works equivalently in the Lagrangian path integral framework, where we deal with integrals
\begin{equation}
\int D\phi\exp\int\rmd^4x \rmi\mathcal L(\phi(x),\partial\phi(x),...)
\end{equation}
with the local form of $\mathcal L$ and field $\phi$. Then the local choice can be realized by adding $\mathcal L\to\mathcal L+a(x)V(\phi(x))$. Relativistic invariant choice means no changes of choice-dependent $\mathcal L$ under Lorentz transformations, applied to both $a$ and $V$. We can take
$V=\phi$ for a scalar field and $a\to a^\mu$, $V\to j^\mu$ or $A^\mu$ in the case of quantum electrodynamics, with current $j$ and potential $A$. 

\subsection{Operational invariance}

According to the Wightman axiom \cite{wight}, a relativistic-invariant Lagrangian should imply invariant quantum correlations of the form
\begin{equation}
\langle \hat{O}_1(x)\hat{O}_2(y)\hat{O}_3(z)\cdots\rangle,
\end{equation}
where the average is defined as $\langle\hat{X}\rangle=\mathrm{Tr}\hat{X}\hat{\rho}$, in the normalized, Hermitian,  and positive definite state $\hat{\rho}$ ($=|\psi\rangle\langle\psi|$ for a pure state). Invariance requires Lorentz transformation of
all $\hat{O}$'s and $\hat{\rho}$.  For free choices the invariance axiom  extends to
\begin{equation}
\langle \hat{O}_{1a}(x)\hat{O}_{2b}(y)\hat{O}_{3c}(z)\cdots\rangle\label{frein}
\end{equation}
The axiom of invariance is not straightforward to prove in general, except free theories.
For interacting theories only in vacuum at zero temperature and perturbatively it has been shown in detail elsewhere  \cite{ab13}. 
Finite temperature states are certainly not invariant themselves which makes the analysis quite hard. Nevertheless, for our purposes the perturbative case of zero temperature is sufficient so we can take the operational invariance for granted.

\section{REALISM AND RELATIVITY}

Realism means a construction of observations described by a set of random functions $o_i(x)$. In the usual quantum mechanics the probability is given by positive operator-valued measure (POVM) \cite{povm}, as
$\langle\hat{K}^\dag\hat{K}\rangle$ with the set of Kraus operators $\hat{K}$ \cite{kraus}.
 The use of POVM is here both ambiguous and obscure, because no single POVM can be reliably distinguished and even if we determine one any calculations will be tedious. Even worse, every POVM (even apparently those that are invariant with respect to relativity) makes the dynamics disturbed and is irreversible, which is a common problem of objective collapse theories \cite{grw,csl}. Here we do not accept such a disturbance in objective realism and demand strict noninvasiveness of observations.   Irreversibility is still possible due to largeness and openness of the system but not the observations themselves.
A better approach requires the framework of weak measurements \cite{weak} which are a special limit of a POVM corresponding to a weakly disturbing observation, so that invasiveness disappears in the limit\cite{bb10,abn,bfb}. The price to pay is a large additional Gaussian noise convoluted with the internal statistics. The latter alone must be described by quasiprobability $Q$ (sometimes negative, like the Wigner function \cite{wigner}, in contrast to normal probability)  so it is alone insufficient for realism. In standard quantum measurement theory \cite{povm}, any measurement of finite strength, even weak, leads to some (although tiny) disturbance. On the other hand, the only perfectly nondisturbing standard quantum measurement is trivial -- not measuring anything at all. Therefore, to define noninvasive observations and realism, we have to make a step beyond standard measurement. Namely, we take $Q$ obtained from the noninvasive limit and convolute some extra noise $N$ (but finite) to lift the negativity, which is possible within the experimental regime, discussed in detail in \cite{ab15}. The advantage of such a step is that no collapse is necessary at all, while the noise $N$ reduces the observations to standard projections for sufficiently macroscopic observation (when the noise $N$ becomes irrelevant). In this way we stay as close to standard measurement as possible, yet preserve noninvasiveness. This is consistent, e.g., with the condensed matter approach to quantum noise \cite{clerk}. The real probability $P$ of an observable $o$ localized in spacetime and choice dependent is expected in the form
\begin{equation}
P[o]=N\ast Q=\int \mathrm{D}o'  N[o']Q(o-o'),\label{conv}
\end{equation}
where $N$ is an external noise (positive probability) and $Q$ is an internal quasiprobability. 
The main point of this work is to check if such a construction is possible to include free choice. Namely, all readouts will be choice conditioned, e.g. $o\to o_a,o_{ab}$. This means that readouts for all choices, also those not just realized, are measurable. One can extend this idea naturally to continuous fields and choices, and then $\alpha[x,a]$ is a function of $x$ and functional of $a$. We assume that $N$ is an independent choice and state of the system. Otherwise we would have  additional choice-controlled dynamics. In that case we will rather incorporate all such dependence in the quantum description alone. This is a reasonable minimalist approach, where quantum mechanics essentially captures all the dynamics.

The quasiprobability statistics can be conveniently written in the form for correlations \cite{bfb,bbrb},
\begin{eqnarray}
&&\langle o_1(x_1)\cdots o_n(x_n)\rangle_Q=\nonumber\\
&&\int \mathrm{d}^nx'\:\mathcal T\langle\check{O}_n^{x_n-x'_n}(x'_n)
\cdots\check{O}_1^{x_1-x'_1}(x'_1)\rangle.
\label{avgabs}
\end{eqnarray}
where $\mathcal T$ denotes time ordering, with respect to $x^{\prime 0}$, and
\begin{equation}
\check{O}^{x-x'}(x')
=\delta(x-x')\check{O}^c(x')+f(x-x')\check{O}^q(x')/2\:.\label{avg2}
\end{equation}
The superoperators $\check{O}^{c/q}$ \cite{zwanzig} act on any
operator $\hat{X}$ as an anticommutator/commutator:
$\check{O}^c\hat{X}=\{\hat{O},\hat{X}\}/2$ and
$\rmi\check{O}^q\hat{X}=[\hat{O},\hat{X}]$.  Alternatively $2\check{O}^c=\check{O}^++\check{O}^-$ and $\rmi\check{O}^q=\check{O}^+-\check{O}^-$ with $\check{O}^+\hat{X}=\hat{O}\hat{X}$ and $\check{O}^-\hat{X}=\hat{X}\hat{O}$.
The function $f$ is in principle arbitrary but it turns out that only two choices are reasonable, in particular $f=0$ (no memory)  \cite{bb10,bfb} or
$f(x)=\delta^3(\vec{x})/\pi x^0$ (no correlations in zero temperature equilibrium) \cite{bbrb}.
The operators $\hat{O}$ are given in the Heisenberg picture including the free part governed by the field $a$. In principle in (\ref{avgabs})
one could define correlations for different $a$ and $a'$ (or more) but they are not directly measurable. For our goal it is sufficient to consider a single $a$.

\begin{figure}
\includegraphics[scale=.5]{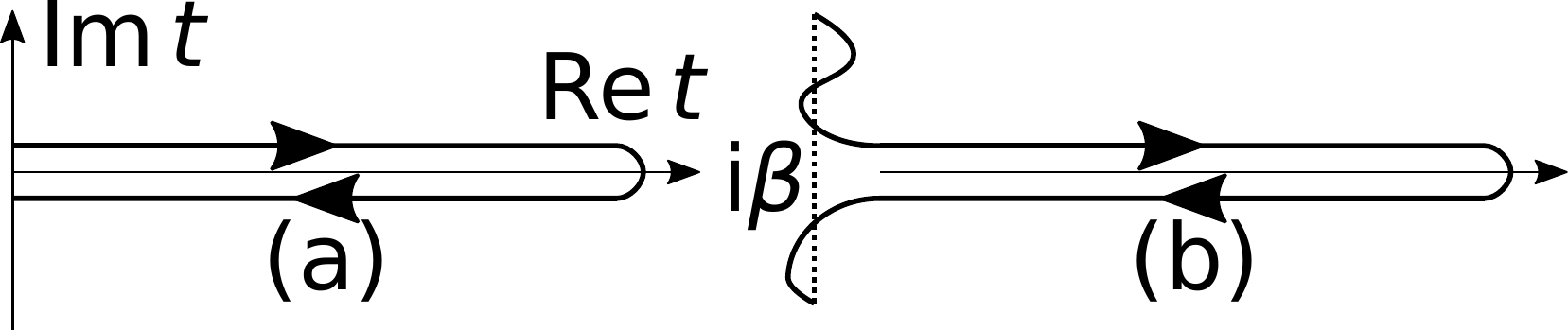}
\caption{Schwinger-Keldysh time contour (a) in general with the left part unspecified and (b) for a thermal state with $\beta=1/k_BT$. The shape of the left line is arbitrary. The time window for observations is bounded by the horizontal part}\label{kel}
\end{figure}

In quantum field theory, the above can be written in terms of path integrals, namely
\begin{eqnarray}
&&\langle X\rangle\int D\phi\exp\int \rmi\rmd^4x \mathcal L(\phi,\partial\phi)\nonumber\\
&&=\int D\phi\exp\int \rmi\rmd^4x \mathcal L(\phi,\partial\phi)X
\end{eqnarray}
with Lagrangian density $\mathcal L$ and integration over $x^0$ along the Schwinger-Keldysh-Kadanoff-Baym contour \cite{ctp0,ctp}, shown in Fig. \ref{kel}(a), where the state is described by properly defining $\mathcal L$ and the path of $x^0$ before the earliest $x^0$ with an active choice or observation.
For instance a thermal state of temperature $T$ means simply extending $x^0$ to complex values with a jump of $\rmi\beta$ ($\beta=1/k_BT$, becomes $\rmi\infty$ at $T\to 0_+$) as shown in Fig.\ref{kel}(b). It is important to discriminate between forward, $+\rmi\rm\epsilon$0 , and backward, $-\rmi\epsilon$, times $x^0_\pm$, respectively, with $\epsilon\to 0_+$ (the spatial position is unaffected) . In such description $\check{O}^\pm(x)\to O(x_\pm)$ and time order is dropped (except the fact that fermion fields are anticommuting Grassmann numbers). Free field $a(x_\pm)=a(x)$ is the same for forward and backward time.

To proceed with the problem of relativistic invariant realism, we have to recall the relativistic framework.
We shall use standard relativistic quantum field notation with four-vectors $A^\mu$ (e.g., field); $x^\mu$ (position in
spacetime); a flat metric $g^{\mu\nu}=g_{\mu\nu}=\mathrm{diag}(1,-1,-1,-1)$; summation convention and 
index shifting
$X\cdot Y=X^\mu Y_\mu=\sum_\mu X^\mu Y_\mu=X^\mu g_{\mu\nu}Y^\nu=X_\mu g^{\mu\nu}Y_\nu$, 
$X^\mu=g^{\mu\nu}X_\nu$, $X_\mu=g_{\mu\nu}X^\nu$, with
derivatives $\partial_\mu=\partial/\partial x^\mu$. Along the Schwinger-Keldysh contour we parametrize $x^0(s)$ by real $s$ with $\rmd x^0=(\rmd x^0/\rmd s)\rmd s$ and \newline $\partial_0=(\rmd x^0/\rmd s)^{-1}\partial_s$. We shall often switch to momentum or Fourier space with $X(p)=\int\rmd^4x\rme^{\rmi p\cdot x}X(x)$,
which needs us to specify $x$ along either the $+$ or $-$ part.
Then the equilibrium $f$  gives Fourier transform $f(p)=\rmi\;\mathrm{sgn}p^0$ \cite{bbrb}.
However, if we want relativistic invariance, the proper choice is $f(p)=\rmi\;\mathrm{sgn}p^0\theta(p\cdot p)$ \cite{ab15}. In any reasonable choice we have $f(p)=0$ for spacelike $p$, $p\cdot p<0$. It essentially means $\check{O}(p)\to (O_+(p)+O_-(p))/2$ for spacelike $p$ or $f=0$
and $\check{O}(p)\to O_\pm(p)$ for the other $f$ and timelike $p$, $p\cdot p>0$, with $\pm p^0>0$.
The invariance of $Q$ follows then from the Wightman axiom (\ref{frein}), so it remains to check if $N$ also can be invariant.

\section{AN ATTEMPT OF INVARIANT REALISM}

A simple convolution with positive $N$ makes it impossible to construct relativistic invariant realism even without free choice, because of zero-temperature counterexamples \cite{ab15}. However we can avoid the zero-temperature problems by simply subtracting zero-temperature statistics. It can be achieved in the following way. The convolution (\ref{conv}) is equivalent to a simple sum of generating functions, namely,
\begin{eqnarray}
&&S_P[\chi]=S_N[\chi]+S_Q[\chi],\nonumber\\
&&\rme^{S_X[\chi]}=\int Do X[o]\exp\int\rmi\rmd^4x o(x)\chi(x).\label{choice}
\end{eqnarray}
Generating functions can be used as a formal series with cumulant expansion in $\chi$, e.g.,
\begin{eqnarray}
&&S(\chi_1,\chi_2)=\rmi \chi_1C_{10}+\rmi \chi_2C_{01}\\
&&-\chi_1^2C_{20}/2-\chi_1\chi_2C_{11}-\chi_2^2C_{02}/2+\dots\nonumber
\end{eqnarray}
with cumulants $C_{10}=\langle o_1\rangle$, $C_{01}=\langle\chi_2\rangle$, $C_{20}=\langle (\delta o_1)^2\rangle$,
$C_{02}=\langle(\delta o_2)^2\rangle$, $C_{11}=\langle\delta o_1\delta o_2\rangle$, $\delta o=o-\langle o\rangle$, etc. There is a one-to-one correspondence between cumulants e.g. $C_{ijk}$ and moments $M_{ijk}=\langle o_1^i o_2^j o_3^k\rangle$ up to a given $i+j+k$, the order of cumulants/moments. 

We assume that only cumulants/moments up to a given order are interesting. It is reasonable because
(a) high order cumulants/moments correspond to low experimental accuracy and complicated unreliable theoretical predictions and (b) for almost all practical purposes (both high and low energy physics) it is sufficient to consider only low order moments (also in tests of locality or contextuality \cite{bbb}). Instead of the full form of $N$ we can only take $S_N$ and even split into some pieces, e.g. $\sum_k S_{N_k}$. For any positive probability the second cumulant $C_{20}$ must be positive. However, this is only necessary only for the sum of all pieces, including $S_Q$. For sufficiently large second order cumulants (correlations), a real positive probability $P$ can be constructed when the cumulants are known up to a given order \cite{ab15}. Therefore we can postulate the arbitrary forms of $S_{N_k}$, as long as the overall $S$ corresponds to a positive probability, in particular second order correlations. 

\subsection{Problem of zero-point correlations}

To show that the construction of objective realism cannot be at all straightforward, let us repeat the conflict caused by zero-point correlations \cite{ab15}. In quantum electrodynamics vacuum current-current correlation must take the form
$\langle j^\mu(p)j^\nu(q)\rangle=(2\pi)^4\delta^4(p+q)G^{\mu\nu}(p)$, where the function $G$ must be positive and invariant so it must be of the form $p^\mu p^\nu\xi+g^{\mu\nu}\eta$ and both $\xi$ and $\eta$ depend only on $p\cdot p$. Positivity leads to $0>(p\cdot p)\eta >-\xi$ for $p\cdot p>0$ and $\eta=0$, $\xi>0$ for $p\cdot p<0$. However, one can find nonzero correlations involving $j(p)$ for $p\cdot p<0$, while $j\cdot p=0$ and the other product of observables $A$, violating Cauchy Schwarz inequality $\langle j(p)j(-p)\rangle\langle|A|^2\rangle\geq|\langle j(p)A\rangle|^2$. Even the scalar field correlation $\langle \phi(p)\phi(q)\rangle=(2\pi)^2\delta(p+q)G(p)$ must be zero if we apply the fluctuation-dissipation theorem \cite{fdt}, leading to analogous violation. To resolve this conflict we 
take one particular piece $S_{N_0}=-S_{Q,vac}$ where $S_{Q,vac}$ is the quantum generating function of the zero-temperature vacuum. This will get rid of any zero-temperature counterexamples because we get null statistics $o=0$ at $T=0$. We shall see later, however, that the vanishing of correlations for spacelike $p$ cannot be resolved if we include freedom of choice.

\subsection{Nonzero temperatures}
Certain problems arise at nonzero temperature, since the correlation function $G$ must be positive. It will be indeed true for $p\cdot p<0$ (spacelike), because the vacuum contribution vanishes and the nonzero-temperature one must be positive. However, for an electron of the mass $m$ and $p\cdot p>m^2$ (timelike) we shall find a negative contribution.
We have $j^\mu(p)j^\nu(q)\to j^\mu_+(p)j^\nu_-(q)$ for $p^0>0$ in the case of $f=\rmi\;\mathrm{sgn}p^0\theta(p\cdot p)$  and $j^\mu(p)j^\nu(q)\to 
(j^\mu_+(p)+j^\mu_-(p))(j^\nu_+(q)+j^\nu_-(q))/4$ for $f=0$. Due to unitarity, we have $\langle\check{X}^q \check{Y}^q\rangle=0$ for 
every $X$ and $Y$ which means that we can subtract $(j^\mu_+(p)-j^\mu_-(p))(j^\mu_+(q)-j^\nu_-(q))/4$ to get 
$(j^\mu_+(p)j^\nu_-(q)+j^\mu_-(p)j^\nu_+(q))/2$ for $f=0$. In terms of fields $j^\mu=\bar{\psi}\gamma^\mu\psi$ with $4\times 4$ Dirac 
matrix $\gamma$ ($\gamma^\mu\gamma^\nu+\gamma^\nu\gamma^\mu=2g^{\mu\nu}$) and Grassmann (anticommuting) fields $\psi$ and 
$\bar\psi=\psi^\dag\gamma^0$. By the standard methods \cite{peskin,ctp,ab13}
\begin{eqnarray}
&&\langle j^\mu_+(p)j^\nu_-(q)\rangle=-(2\pi)^6\delta(p+q)\int\rmd^4k\times\nonumber\\
&&\delta((k+p/2)\cdot(k+p/2)-m^2)\times\nonumber\\
&&\delta((k-p/2)\cdot(k-p/2)-m^2)\times\nonumber\\
&&\left(\frac{\theta(-k^0-p^0/2)}{1+\rme^{-\beta|k^0+p^0/2|}}-\frac{\theta(k^0+p^0/2)}{1+\rme^{\beta|k^0+p^0/2|}}\right)\times\nonumber\\
&&\left(\frac{\theta(k^0-p^0/2)}{1+\rme^{-\beta|k^0-p^0/2|}}-\frac{\theta(p^0/2-k^0)}{1+\rme^{\beta|k^0-p^0/2|}}\right)\times\nonumber\\
&&\mathrm{Tr}\gamma^\mu(\gamma\cdot(k+p/2)+m)\gamma^\nu(\gamma\cdot (k-p/2)+m).
\end{eqnarray}
Evaluating the trace (last line) gives $8k^\mu k^\nu-2p^\mu p^\nu-g^{\mu\nu}(4k\cdot k-p\cdot p-4m^2)$. 
Combining $(k\pm p/2)\cdot(k\pm p/2)=m^2$ we get additionally $k\cdot p=0$ and $k\cdot k+p\cdot p/4=m^2$, so the trace becomes 
$8k^\mu k^\nu+2(g^{\mu\nu}p\cdot p-p^\mu p^\nu)$. 
The difference between finite and zero temperature has the form
\begin{eqnarray}
&&\langle j^\mu_+(p)j^\nu_-(q)\rangle_{T-0}=(2\pi)^6\int\rmd^4k\times\nonumber\\
&&\delta(k\cdot p)\delta(k\cdot k+p\cdot p/4-m^2)\nonumber\\
&&\left[(1+\rme^{\beta|k^0+p^0/2|})^{-1}
(1+\rme^{\beta|k^0-p^0/2|})^{-1}\vphantom{frac{\theta k^0}{\rme^{\beta k^0}}}\right.\nonumber\\
&&\left.
-\frac{\theta(-k^0-p^0/2)}{1+\rme^{\beta|k^0-p^0/2|}}
-\frac{\theta(k^0-p^0/2)}{1+\rme^{\beta|k^0+p^0/2|}}\right]\nonumber\\
&&\times(2(p^\mu p^\nu-g^{\mu\nu} p\cdot p)-8k^\mu k^\nu).\label{corx}
\end{eqnarray}
The last line is positive definite for timelike $p$ and negative definite for spacelike $p$.
For the timelike case, let us take the frame where $p=(P,0,0,0)$ and then $k=(0,K,0,0)$, and $K^2=P^2/4-m^2$. Then we get only nonzero elements $8m^2$ for $\mu=\nu=1$ and $2P^2$ for $\mu=\nu=2,3$. For the spacelike case we take $p=(0,P,0,0)$ so $k=(K_0,0,K,0)$
with $K_0^2=m^2+P^2/4+K^2$. The only nonzero elements are $-8(K^2+m^2)$ for $\mu=\nu=0$, $-2P^2-8K^2$ for $\mu=\nu=2$, $-8K_0K$ for $\mu\nu=20,02$ and $-2P^2$ for $\mu=\nu=3$. The negativity is confirmed by the Cauchy-Schwarz inequality $(P^2+4K^2)(K^2+m^2)- 4K_0^2K^2=P^2m^2\geq 0$.

Now, the middle line in (\ref{corx}) is always negative. This is because either $p^0>0$ which leaves only one $\theta$ while all 
Fermi factors $(1+\rme^{\beta q})^{-1}<1$ or we symmetrize contributions from $p$ and $q=-p$, which turns both $\theta$ into $1/2$ and the same argument applies. Therefore $G$ is positive definite for spacelike $p$ but negative definite for timelike $p$ with
 $p\cdot p>4m^2$ and zero for $4m^2>p\cdot p>0$. To fix the problem of  positivity we need to add another $S_{N_1}$ with positive definite correlation for $p\cdot p>0$. To this end, we can take, e.g., the bosonic Proca field $B^\mu(x)$ with the Lagrangian
 $2\mathcal L=B_{\mu\nu}B^{\nu\mu}+M^2B\cdot B+\xi(\partial\cdot B)^2$ with $B_{\mu\nu}=\partial_\mu B_\nu-\partial_\nu B_\mu$
 and $\xi\to+\infty$ (Lorentz gauge fixing $\partial\cdot B=0$). Then 
\begin{eqnarray}
&&\langle B^\mu_+(p) B^\nu_-(q)\rangle=\nonumber\\
&&(2\pi)^5\delta(p+q)(p^\mu p^\nu-g^{\mu\nu}p\cdot p)
 \delta(p\cdot p-M^2)\times\nonumber\\
&&\left(\frac{\theta(-p^0)}{\rme^{\beta|p^0|}-1}+\frac{\theta(p^0)}{1-\rme^{-\beta|p^0|}}\right).
\end{eqnarray}
 We can now redefine the observable current \newline $j^\mu\to j^\mu+\int \rmd M \eta_M B^\mu_M$ with some form factor $\eta$. Alternatively, we can take an abstract field $B^\mu$ with the correlation $\langle B^\mu(p)B^\nu(q)\rangle=(2\pi)^4\delta(p+q)(p^\mu p^\nu-g^{\mu\nu}p\cdot p)X(p\cdot p)$ with some positive function $X$, which is zero for negative arguments.
 
 For a maximally spacelike case in (\ref{corx}), $p^0=0$, the middle line reads $-(2\cosh(\beta|K_0|/2)^{-2}$, while 
$K_0^2>m^2+P^2/4$. At low temperatures (large $\beta$) it vanishes exponentially at least $\sim \rme^{-\beta m}$, but the same behavior applies to all correlation functions. Therefore we cannot construct (at least easily) an example against realism in this case, because of the positivity of second order correlations, without freedom of choice.

\section{FAILURE OF INVARIANT FREE CHOICE}

Now we will show that relativistic invariant realism breaks down when we introduce freedom of choice. Let us add a free part to the Lagrangian density (at some point $x$) of either the scalar field $\phi$ or electron spinor $\psi$,
\begin{eqnarray}
&&2\mathcal L=(\partial\phi)\cdot(\partial\phi)-m^2\phi^2+\lambda\phi^4/12+2a\phi,\nonumber\\
&&\mathcal L=\bar\psi(\rmi\gamma\cdot\partial-m+\gamma\cdot A)\psi,\label{fre}
\end{eqnarray}
where $a$ and $A$ are freely chosen external fields. Here $\lambda$ introduces nonlinear interaction because the linear scalar case is trivial and agrees with realism, so the distribution at $a=0$ will be simply shifted by $\phi\to\phi+a$.
All correlations in $S_Q$ start to depend on $a$ or $A$ but not those in $S_N$ in (\ref{choice}) as the choice applies only to the standard quantum part. The invariance condition is that, in the limit of zero temperature, they stay invariant under simultaneous change of the frame for $\phi$, $a$, $j=\bar{\psi}\gamma\psi$ and $A$, according to Lorentz rules.
To show that this is impossible, we take $a$ and $A$ as small parameters and expand all correlations in their powers, e.g.,
\begin{eqnarray}
&&\langle \phi(x)\phi(y)\rangle=G_0(x,y)+\int \rmd^4z G_1(x,y;-z)a(z)\nonumber\\
&&+\int\rmd^4 z\rmd^4 w G_2(x,y;-z,-w)a(z)a(w)+...
\end{eqnarray}
Certainly $G_0$ corresponds to the zero-temperature vacuum limit of the previous case. We have already learned that $G_0(p,q)=0$ for spacelike $p$ (or $q$). We assume that $a$ is sufficiently small so that we can perform perturbative analysis, comparing correlations expanded to the same maximal power of $a$.
Let us consider the function $\langle\phi(p)\phi(q)\phi(k)\rangle$ in equilibrium vacuum for spacelike $p$, $q$, $k$ and any sum of them. Then $2\phi\to \phi_++\phi_-$
and
\begin{equation}
\langle\phi(p)\phi(q)\phi(k)\rangle=\int \rmd^4 s\langle \phi^c(p)\phi^c(q)\phi^c(k)\phi^q(s)\rangle b(s).\label{pp1}
\end{equation}
We shall focus on the expression $\langle \phi^c(p)\phi^c(q)\phi^c(k)\phi^q(s)\rangle$ (also called susceptibility). From unitarity we can add 
$\langle \phi^q(p)\phi^q(q)\phi^q(k)\phi^q(s)\rangle/8\rmi$ (which is zero). We shall obtain various combinations of the Schwinger-Keldysh parts of the contour ($+$ or $-$), but in particular there will be
$++++$  but not $----$ [because crossings $+-$ or $-+$ must be timelike; see also (\ref{fdt1}) and the discussion below]. The expectations will also contain $\delta(p+q+k+s)$. We can take, e.g., vertices of the regular tetrahedron, $p^0=q^0=k^0=s^0=0$ and $\vec{p}=C(1,1,1)$, $\vec{q}=C(1,-1,-1)$, $\vec{k}=C(-1,1,-1)$, and $\vec{s}=(-1,-1,1)$. Then only the term $++++$ will contribute, which is at zero temperature
\begin{equation}
\lambda[(p\cdot p-m^2)(q\cdot q-m^2)(k\cdot k^2-m^2)(s\cdot s-m^2)]^{-1},
\end{equation}
which is $\lambda/(m^2+3C^2)^4$ for the tetrahedron.
On the other hand, realism requires the Cauchy-Schwarz inequality
\begin{equation}
|\langle\phi(p)\phi(q)\phi(k)\rangle|^2\leq \langle|\phi(p)|^2\rangle\langle|\phi(q)\phi(k)|^2\rangle\label{csch}
\end{equation}
with regularization $\phi(w)\to \int \rmd^4v \delta_\epsilon(v-w)\phi(v)$.
However, the left-hand side is nonzero and proportional to $\lambda^2 |b(s)|^2$ while on the right-hand side 
$\langle|\phi(p)|^2\rangle$ disappears if, for $a(s)$,  $p+ns$ is spacelike for all integer $n$ and the inequality is obviously violated in zero temperature vacuum. Note that the example has no proper classical limit, at least at zero temperature. This is because time-resolved observation is burdened with time-frequency uncertainty and even the simple vacuum fluctuations (zero-point quantum noise) do not contain the Planck constant (tracing back the dimension) and the only comparison scale is the mass of a (charged) particle , which is combined with the Planck constant and speed of light to get the frequency dimension.

It is interesting to understand why there is no contribution from $a(s)$. Let us expand
\begin{equation}
\langle\phi(p)\phi(p')\rangle=\int\rmd^4 s\rmd^4 s' G_2(p,p',s,s')a(s)a(s')+...\label{ppp}
\end{equation}
The zero order term vanishes because $p,p'$ are spacelike and because of arguments analogous to those in \cite{ab15}, repeated here in Sec. IVA and the first order one from parity.
The remaining $G_2$ corresponds to \newline $\langle \phi^c(p)\phi^c(p')\phi^q(s)\phi^q(s')\rangle$. From unitarity we add \newline
$\langle\phi^q\phi^q\phi^q\phi^q\rangle/4$, which leaves only the terms $+-\ast\ast$ and $-+\ast\ast$, so $p$ and $p'$ lie on the opposite 
branches of Schwinger-Keldysh contour. They are spacelike, also with added $s,s'$, so there is no possibility to go between branches -- there is always  $\delta_+(w\cdot w-m^2)$ from $+$ to $-$ so the sum of all transfer variables $w$'s, must be timelike but also equal $p$, $p+s$, or $p+s'$, which is a
contradiction. The argument extends analogously to higher orders  with the restriction that the sum  $p+\sum_i s_i$ cannot become timelike. However, instead of showing that (\ref{ppp}) vanishes, it is sufficient to show that it is at least $\sim |a|^4$.

\begin{figure}
\includegraphics[scale=.7]{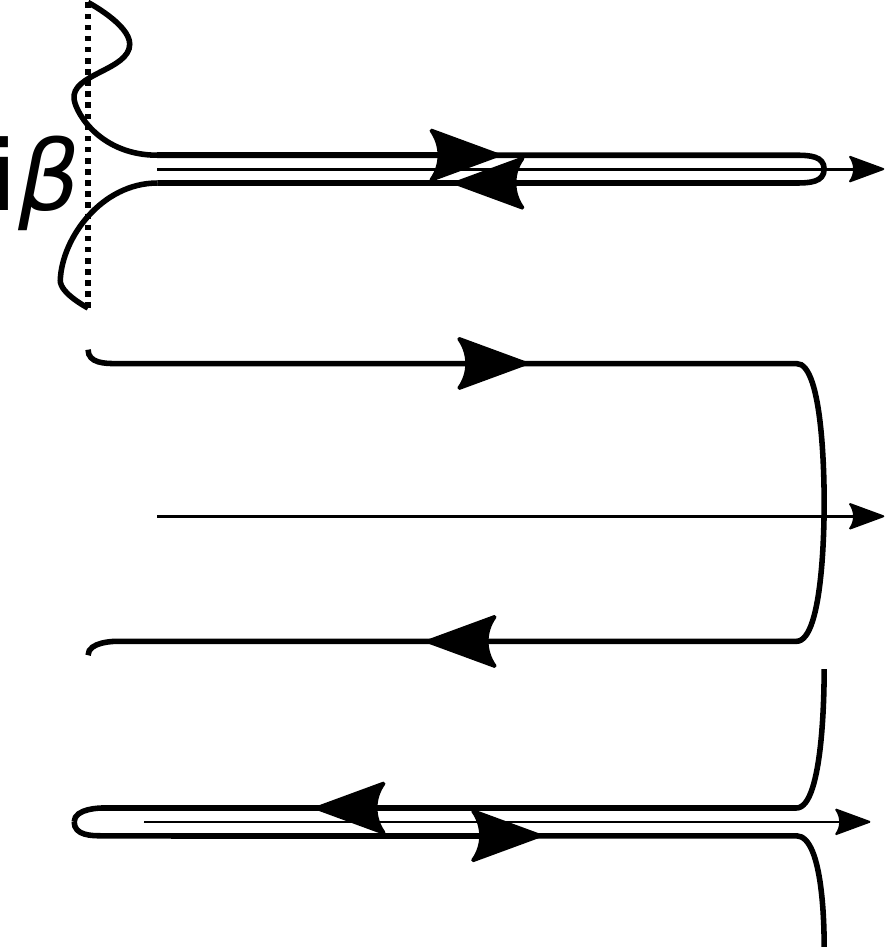}
\caption{Transformation of the Schwinger-Keldysh-Kadanoff-Baym contour leading to the generalized fluctuation-dissipation theorem (\ref{fdt1}). In the first stage the flat parts are moved away; in the second stage the contour is cut on the right and glued on the left.}
\label{fdt}
\end{figure}

Alternatively, we can use a generalized form of quantum fluctuation theorem for thermal states \cite{fdt}, namely,
\begin{eqnarray}
&&\left\langle \prod_i O_{i-}(p_i)\prod_jO_{j+}(p_j)\right\rangle\exp\sum_j\beta p^0_j=\nonumber\\
&&\left\langle \prod_j O_{j-}(p_j)\prod_iO_{i+}(p_i)\right\rangle^\ast_{r},
\label{fdt1}
\end{eqnarray}
where $r$ denotes the time reversal of fields and of the Lagrangian. Here $\phi_r= \phi$, $X^\mu_r=(-1)^\mu X^\mu$, with $(-1)^0=1$ and \newline $(-1)^{1,2,3}=-1$ for $X=A,B,p,j$. It can be easily proved by modifying the Schwinger-Keldysh-Kadanoff-Baym contour as shown in Fig. \ref{fdt} where we separate the horizontal part by $\rmi\beta$, which results in additional factors $\rme^{\beta p^0_j}$. Note also that $\sum_ip^0_i+\sum_jp^0_j=0$ because of time shift invariance.
In the last step we have to reverse time, which is accompanied by conjugation because time reversal is antiunitary.
Now, in the zero-temperature limit averages are relativistic invariant but also the exponent $\sum_j \beta p^0_j$ diverges unless
$\sum_j p^0_j=0$. Therefore these averages must vanish if $\sum_j p_j$ is spacelike because we can find a frame where $\sum_j p^0_j\sim 0$, i.e. minimal changes will reverse the sign. For timelike $\sum_j p_j$ the average on the right-hand side of (\ref{fdt1}) must vanish if $\sum_j p^0_j<0$.

An analogous example involves current, namely, \newline $\langle j^\mu(p)j^\nu(q)j^\sigma(k)\rangle$ at free choice $A^\tau(s)$.
Then (\ref{pp1}) for spacelike $p,q,k,s$ takes the form
\begin{equation}
\int\rmd^4s \langle j^\mu_+(p)j^\nu_+(q)j^\sigma_+(k)j^\tau_+(s)\rangle A_\tau(s),
\end{equation}
which is a four-point Green function discussed a long time ago \cite{g44}. We recall the calculation in the Appendix with the lowest order limit, for
$p,q,k,s\ll m$. Let us take $\mu=\nu=\sigma=\tau=0$ and again vertices of regular tetrahedron. Then
\begin{eqnarray}
&&\int\rmd^4s \langle j^0_+(p)j^0_+(q)j^0_+(k)j^0_+(s)\rangle=\nonumber\\
&&-(2\pi)^4\delta(p+q+k+s)
16\pi^2 C^4/15m^4,
\end{eqnarray}
which is clearly nonzero, contradicting an analogue of (\ref{csch}) with $\phi\to j^0$ because $\langle j^\mu(p)j^\nu(-p)\rangle$ will be zero (or $\sim |A|^4$; the zeroth order vanishes as shown in \cite{ab15} and Sec. IVA).

We have shown that an attempt to build free choice into quantum mechanics fails when trying to reconcile with relativity.
If we abandon relativistic invariance we can make $\langle|\phi(p)|^2\rangle$ positive for every $p$, not only timelike. The failure is generic as it occurs both for scalar and vector (spinor) fields. 

\section{RELATIVISTIC INVARIANCE AND NO-SIGNALING}

One of the consequences of relativistic invariance is the principle of no-signaling. It states that the correlations \newline 
$\langle\prod_j\phi_i(x_i)\rangle$ cannot depend on free choices $a_j$ localized at $y_j$ so that $x_i-y_j$ is spacelike for all $i,j$. Plainly, 
it forbids superluminal, faster than light, communication. It is justified by the relativistic invariance of correlations because the 
influence associated with $a(y)$ is associated with $\check{\phi}^q(y)$. Because $x-y$ is spacelike, we can find a frame where $y^0=x^0$ 
when  $\hat{\phi}'(x)\hat{\phi}(y)=\hat{\phi}(y)\hat{\phi}'(x)$, 
so $\check{\phi}^q(y)$ gets eliminated. As already stressed, the invariance itself can be 
proved at least perturbatively \cite{ab13} but it is rather accepted as part of Wightman axioms (\ref{frein}), which in fact state both 
invariance (of the vacuum ground state) and no-signaling, also called microcausality \cite{wight}. 
However, once relativistic invariance is put in doubt, no-signaling loses its obvious justification.

\begin{figure}
\includegraphics[scale=.3]{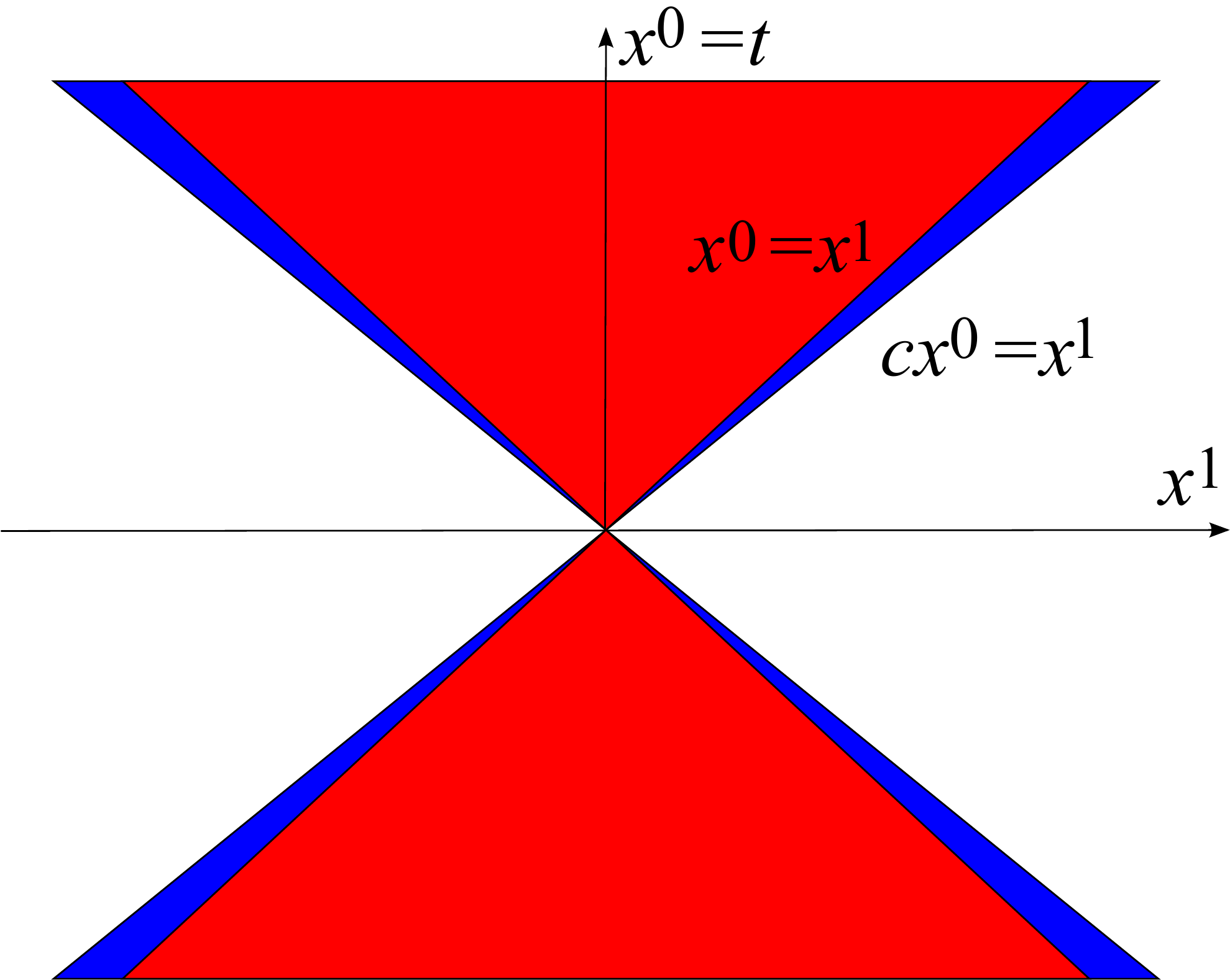}
\caption{Problem of perturbative no-signaling. The exact result gives the signaling speed $c$, bounding the blue area. Starting from  a field with signaling velocity $1$ (bounding the red area) with perturbative expansion the boundary of signaling remains the same (red area).}
\label{ccc}
\end{figure}

One can still ask if adding noninvariant corrections to an invariant theory may lead to the violation of no-signaling. 
We shall demonstrate that indeed it can be violated, but nonperturbatively while the perturbative approach is misleading. Let us look at a counterexample, depicted in Fig. \ref{ccc}.
Let us take a real scalar field $\phi$ with the Lagrangian density analogous to (\ref{fre})
\begin{equation}
2\mathcal L=(\partial_0\phi)^2-c^2(\nabla\phi)^2/2-m^2\phi^2+2b(x)\phi(x).
\end{equation}
It is clear that the signaling speed is $c$ and the causal Green function (commutator) $G(x-y)=\langle\phi^q(x)\phi^c(y)\rangle$ can be written as
\cite{peskin}
\begin{equation}
G(x)=\mathrm{Re}\int\frac{2 d^4q}{(2\pi)^4}\frac{e^{\rmi q_0 x^0-\rmi\vec{q}\cdot\vec{x}}}{(q_0+\rmi\epsilon)^2-c^2|\vec{q}|^2-m^2}
\end{equation}
with $\epsilon\to 0_+$
and can be evaluated exactly as
\begin{eqnarray}
&&\frac{m\theta(c|x^0|-|\vec{x}|)}{4\pi c^2}\frac{J_1(m\sqrt{|x^0|^2-|\vec{x}/c|^2})}{\sqrt{|cx^0|^2-|\vec{x}|^2}}\nonumber\\
&&
-\delta(|cx_0|^2-|\vec{x}|^2)\frac{\mathrm{sgn}{x^0}}{2\pi c},
\end{eqnarray}
where $J$ is the Bessel function. In quantum field theory we need to subtract the renormalizing Green function with a large mass $M^2\gg m^2$, giving effectively
\begin{equation}
G_{r}(x)=\frac{\theta(|cx^0|-|\vec{x}|)}{4\pi c^2}\frac{J_1(mc^2\sqrt{|x^0|^2-|\vec{x}/c|^2})}{\sqrt{|cx^0|^2-|\vec{x}|^2}/m}- m\to M.
\end{equation}
The Green function is not zero only inside the causal cone given by $|\vec{x}|<c|x^0|$, defining the signaling speed as $c$.
Now, let us solve the problem perturbatively, rewriting 
\begin{equation}
c^2(\nabla\phi)^2=(\nabla\phi)^2+\lambda(\nabla\phi)^2,
\end{equation}
where $\lambda=c^2-1$ is a (small) perturbative parameter.
The perturbative solution leads to changing $c$ at constant $x$ and reads
\begin{equation}
G_{r}^{p}(x)=\frac{\theta(|x^0|-|\vec{x}|)}{4\pi c^2}\frac{J_1(m\sqrt{|x^0|^2-|\vec{x}/c|^2}}{\sqrt{|cx^0|^2-|\vec{x}|^2}/m}- m\to M,
\end{equation}
while for negative $\lambda$  and $|x^0|>|\vec{x}|>c|x^0|$ we substitute $J_1(is)=iI_1(s)$, an analytic continuation at $s=0$.
This  is of course different from the exact solution and the reason is that the boundary of the signaling cone limits the validity of perturbative expansion.
The root of the problem is the Fourier representation
\begin{eqnarray}
&&[(q_0+\rmi\epsilon)^2-c^2Q^2-m^2]^{-1}=\nonumber\\
&&[(q_0+\rmi\epsilon)^2-Q^2-m^2-\lambda Q^2]^{-1}=\\
&&[(q_0+\rmi\epsilon)^2-Q^2-m^2]^{-1}+\nonumber\\
&&\lambda Q^2[(q_0+\rmi\epsilon)^2-Q^2-m^2]^{-2}+\nonumber\\
&&\lambda^2 Q^4[(q_0+\rmi\epsilon)^2-Q^2-m^2]^{-3}+\dots\nonumber
\end{eqnarray}
for $Q=|\vec{q}|$. Due to the pole, the geometric series is convergent only at $\lambda Q^2<q_0^2-m^2$, despite leading to a finite contribution at each order of $\lambda$. Beyond the convergence region, summation is only formal and interpreted rather as an analytic continuation.
Therefore, this reasoning is certainly nonperturbative. In principle one could include the analytic continuation of such a series in one of the rules of quantum field theory; all dangerous examples in interacting theories, e.g., bound states and higher order correlation functions, are impossible to check.

We conclude that no-signaling is a nonperturbative principle inherently related to relativistic invariance. This means that
relativistic invariance may be renounced either by a direct experiment in different frames \cite{kost} or indirectly by testing no-signaling.

No-signaling can be simply tested by checking if a free choice can change a spacelike readout. It a necessary assumption of every Bell test \cite{bell,chsh,eber} and therefore it is tested there simultaneously. Although in recent experiments \cite{hanson,zein,saew} the signature of superluminal signaling based on the reported data seems to be yet insignificant, in all of them both random choices and readouts are machine made so fair time tagging and choice is a matter of trust in electronics, not humans -- the choice is not free in the human sense \cite{wise}. Further and improved experiments should be continued to resolve the question of possible superluminal signaling.

\section{CONCLUSIONS}

The presented direct conflicts of freedom of choice in quantum realism with relativity demonstrates incompleteness of the present quantum framework without using the assumption of the Bell theorem. The easiest way seems to abandon relativistic invariance. This can be tested experimentally, especially by no-signaling in the test of local realism, which is different from the direct search for violations of relativistic invariance \cite{kost}. Theoretical and experimental development of such tests is critical for finding a way to reconcile quantum realism with free choices. Finally, the freedom of choice remains a matter of trust in electronics, with human choice yet to be considered \cite{wise}.

\section*{ACKNOWLEDGMENTS}
W. Belzig, R. Demkowicz-Dobrza\'nski, and P. Chankowski are acknowledged for motivation, discussion, and suggestions.

\section*{APPENDIX FOUR-POINT GREEN FUNCTION}

We shall recall the calculation of the four-point electron Green function \cite{g44} defined as follows:
\begin{equation}
G^{\alpha\beta\gamma\delta}(x_ax_bx_cx_d)=\langle j^\alpha_+(x_a)j^\beta_+(x_b)j^\gamma_+(x_c)j^\delta_+(x_d)\rangle.
\end{equation}
We will rather refer to the Fourier transform of $G(a,b,c,d)$ equal to $\langle j_+(a)j_+(b)j_+(c)j_+(d)\rangle$ with
the Fourier transform $j(a)=\int \rmd^4 x j(x_a)\rme^{\rmi x_a\cdot a}$. Thanks to translational invariance $G=\delta(a+b+c+d)\tilde G$.
The calculation of $\tilde{G}$  by standard methods (Wick decomposition into propagators -- two-point Green functions) reduces to three integrals (differing by permutation), corresponding to the box Feynman diagram depicted in Fig. \ref{gabcd}:

\begin{figure}
\includegraphics[scale=.5]{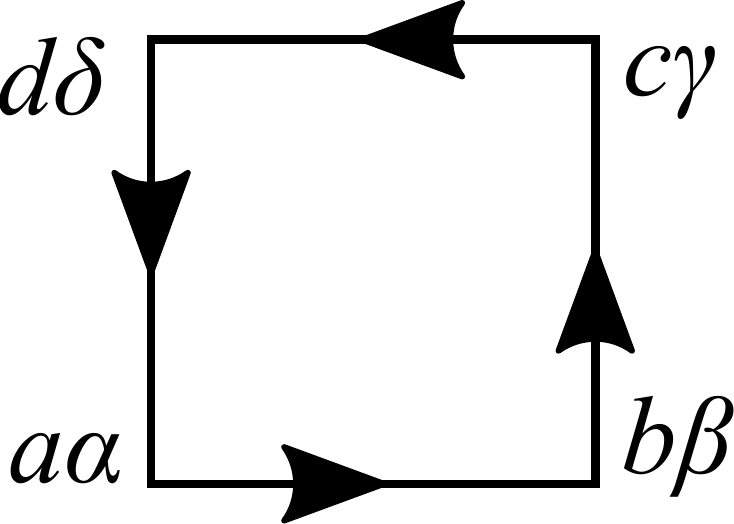}
\caption{One of three (or six when directions are counted) diagrams $T$ contributing to $G$. Gamma matrices are inserted in vertices and propagators in lines}\label{gabcd}
\end{figure}

\begin{equation}
\tilde{G}^{\alpha\beta\gamma\delta}(abcd)=T^{\alpha\beta\gamma\delta}(abcd)+T^{\beta\alpha\gamma\delta}(bacd)
T^{\delta\beta\gamma\alpha}(dbca)
\end{equation} with
\begin{eqnarray}
&&T^{\alpha\beta\gamma\delta}(abcd)=-2\int \rmd^4 p
\mathrm{Tr}(\slashed{p}-m+\rmi\epsilon)^{-1}\gamma^\alpha\times\nonumber\\
&&(\slashed{p}+\slashed{a}-m+\rmi\epsilon)^{-1}\gamma^\beta
(\slashed{p}+\slashed{a}+\slashed{b}-m+\rmi\epsilon)^{-1}\times\nonumber\\
&&\gamma^\gamma(\slashed{p}-\slashed{d}-m+\rmi\epsilon)\gamma^\delta,
\label{ttt}
\end{eqnarray}
where the minus is due to anticommuting, $\slashed{p}=\gamma\cdot p$, the factor $2$ due to opposite directions and $\epsilon\to 0_+$ due to the limits of $x^0$. It is important that $G$ (not $T$) be invariant with respect to the permutation of pairs $(a,\alpha)$, $(b,\beta)$, $(c,\gamma)$, $(d,\delta)$; relativistically invariant; and gauge invariant, namely $a_\alpha G^{\alpha\beta\gamma\delta}(abcd)=0$ (and analogously for other pairs). It is easily proved by the identity
\begin{eqnarray}
&&(\slashed{p}-m+\rmi\epsilon)^{-1}-(\slashed{p}+\slashed{a}-m+\rmi\epsilon)^{-1}=\nonumber\\
&&(\slashed{p}-m+\rmi\epsilon)^{-1}\slashed{a}(\slashed{p}+\slashed{a}-m+\rmi\epsilon)^{-1}
\end{eqnarray}
used at every occurrence of $\slashed{a}$, and by telescoping the cancellation of the left-hand sides from all parts of the integral, with a shift of $p$ when appropriate. 

\begin{figure}
\includegraphics[scale=.45]{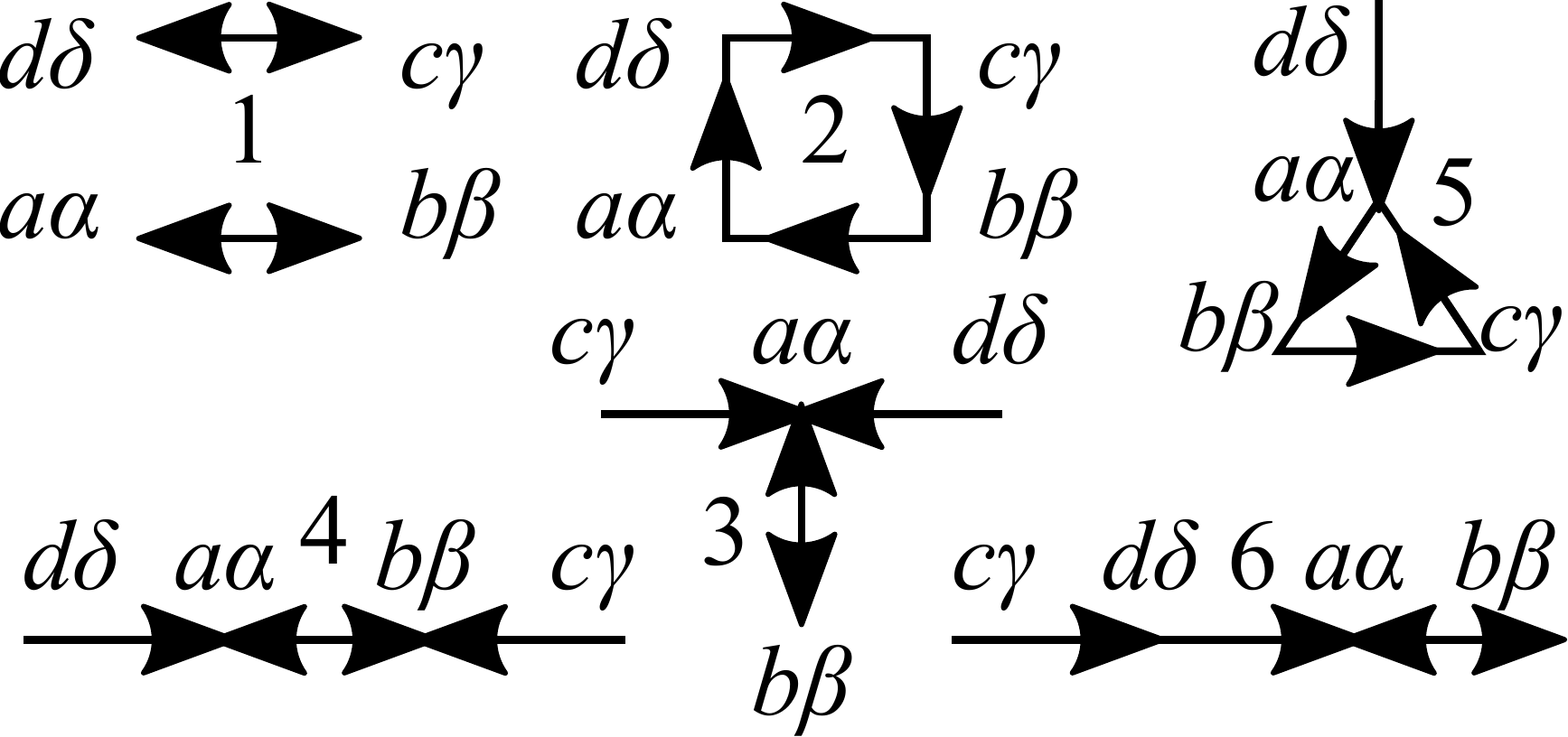}
\caption{All types of heads described in (\ref{hea}). The arrow points from the greek index $\alpha\beta\gamma\delta$ to the latin one $abcd$.}
\label{heads}
\end{figure}

From relativistic invariance $G$ must consist of three types of terms, $k^\alpha q^\beta r^\gamma s^\delta$ (heads), $g^{\alpha\beta}r^\gamma s^\delta$ (and permutations), and $g^{\alpha\beta}g^{\gamma\delta}$ (and two other permutations) multiplied by scalar functions of $abcd$.
Here $kqrs$ are equal to some of $abcd$ but from the condition $a+b+c+d=0$ we can exclude $a$ from $k$, $b$ from $q$, $c$ from $r$, and $d$ from $s$ (by substituting $a=-b-c-d$, etc.). From gauge invariance the heads determine all other terms, because the terms with $g$ cannot exist without heads. It is clear when e.g. we combine $g^{\alpha\beta}r^\gamma s^\delta$ with $a_\alpha$, which gives $a^\beta r^\gamma s^\delta$. Without heads this term can be canceled only by $g^{\alpha\gamma}q^\beta s^\delta$ or $g^{\alpha\delta}q^\beta r^\gamma$.
This is impossible if $r,s\neq a$. By interchanging $\alpha\leftrightarrow \beta$ we find that $rs$ must correspond to $ab$ or $ba$.
However, then $g^{\alpha\gamma}q^\beta s^\delta$ implies $q=c$ and $g^{\alpha\delta}q^\beta r^\gamma$ implies $q=d$, which again makes the cancellation impossible. Terms $g^{\alpha\beta}g^{\gamma\delta}$ left alone have nothing to cancel with. 
Heads can be classified into six types:
\begin{eqnarray}
&&1:\:b^\alpha a^\beta d^\gamma c^\delta,\:2:\:d^\alpha a^\beta b^\gamma c^\delta,\:3:\:b^\alpha a^\beta a^\gamma a^\delta,\label{hea}\\
&&4:\:b^\alpha a^\beta b^\gamma a^\delta,
\:5:\:b^\alpha c^\beta a^\gamma a^\delta,
\:6:\:b^\alpha a^\beta d^\gamma a^\delta,\nonumber
\end{eqnarray}
depicted in Fig. \ref{heads}.

Gauge invariance should allow us to write \newline $H=A_\alpha B_\beta C_\gamma D_\delta G^{\alpha\beta\gamma\delta}$ in terms of fields
$A_{\mu\nu}=\partial_\mu A_\nu-\partial_\nu A_\mu$ or in the Fourier representation $i(a_\nu A_\mu-a_\mu A_\nu)$ (similarly for $BCD$).
In each combination we get some heads
There are plenty of possible types and combinations of fields that may contribute to $H$, but we shall classify five of them (the only relevant ones as we shall see in the end) with corresponding heads in $G$. We shall use compact notation, $k\cdot F\cdot q=k_\mu F^{\mu\nu}q_\nu$ and
$(\cdot X \cdot)=X^\mu_\mu$.
The list is
\begin{eqnarray}
1:&&(\cdot A\cdot B\cdot)(\cdot C\cdot D\cdot)\to  b^\alpha a^\beta d^\gamma c^\delta, \nonumber\\
2:&&(\cdot A\cdot B\cdot C\cdot D\cdot)\to d^\alpha a^\beta b^\gamma c^\delta +b^\alpha c^\beta d^\gamma a^\delta,\label{ten}\\
3:&&(\cdot A\cdot B\cdot)(a\cdot C\cdot D\cdot a)\to \nonumber\\
&&2b^\alpha a^\beta ((a\cdot c)d^\gamma a^\delta+(a\cdot d)a^\gamma c^\delta-(c\cdot d)a^\gamma a^\delta),\nonumber\\
4:&&(\cdot A\cdot B\cdot)(b\cdot C\cdot D\cdot a)\to \nonumber\\
&&2b^\alpha a^\beta((b\cdot c)d^\gamma a^\delta+(a\cdot d)b^\gamma c^\delta-(c\cdot d)b^\gamma a^\delta),\nonumber\\
5:&&(b\cdot D\cdot a)(\cdot A\cdot B\cdot C\cdot)\to\nonumber\\
&&(b^\alpha c^\beta a^\gamma-c^\alpha a^\beta b^\gamma)((b\cdot d)a^\delta-(a\cdot d)b^\delta).\nonumber
\end{eqnarray}
It is clear that the type number matches the head type except for head type 6 which contributes to 3 and 4 here. The tensor 5 is a simplified form of that of \cite{g44} due to the Bianchi identity $c_{\mu}C_{\nu\tau}+c_{\nu}C_{\tau\mu}+c_{\tau}C_{\mu\tau}=0$.

\begin{figure}
\includegraphics[scale=.35]{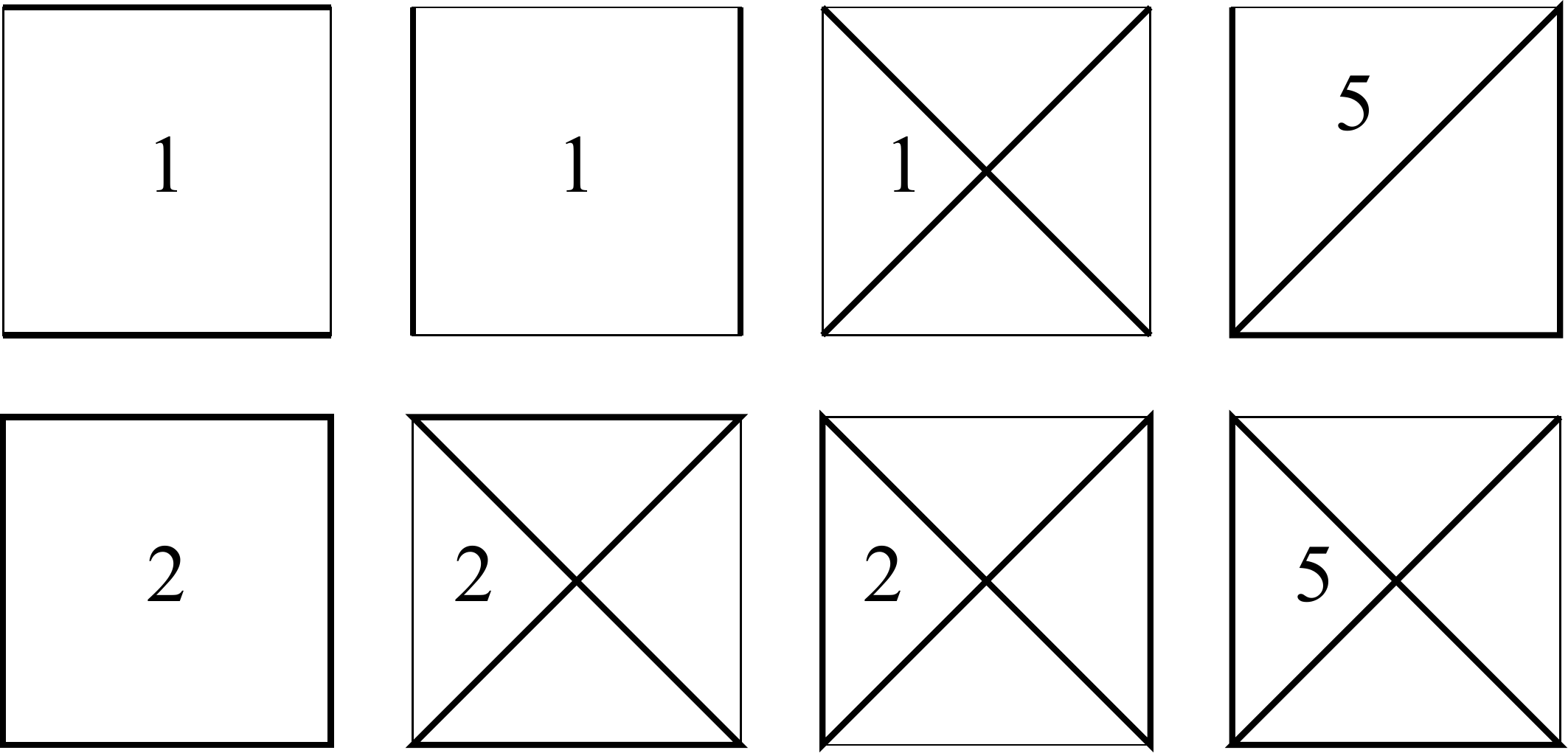}
\caption{Three contributions to the head types 1 and 2 and two to type 5, as in Fig. \ref{heads}, inscribed in the box diagram, Fig. \ref{gabcd}.}
\label{hour}
\end{figure}

To continue the calculation, we rewrite in (\ref{ttt}) 
\begin{equation}
(\slashed{p}-m+i\epsilon)^{-1}=\frac{\slashed{p}+m}{p\cdot p-m^2+\rmi\epsilon}
\end{equation}
and analogously other factors. Then we use the Feynman identity
\begin{eqnarray}
&&(X_{ab}X_{bc}X_{cd}X_{da})^{-1}=\int_0^1 6\rmd^4\lambda\times\nonumber\\
&& \delta(1-\lambda_{ab}-\lambda_{bc}-\lambda_{cd}-
\lambda_{da})\times\nonumber\\
&&(\lambda_{ab}X_{ab}+\lambda_{bc}X_{bc}+\lambda_{cd}X_{cd}+\lambda_{da}X_{da})^{-4}
\end{eqnarray}
applied to $X_{da}=p\cdot p-m^2+\rmi\epsilon$, $X_{ab}=(p+a)\cdot(p+a)-m^2+\rmi\epsilon$, $X_{bc}=(p+a+b)\cdot(p+a+b)-m^2+i\epsilon$, and 
$X_{cd}=(p-d)\cdot(p-d)-m^2+\rmi\epsilon$.
Moreover, we make the shift $p\to p-\lambda_{ab}a-\lambda_{bc}(a+b)+\lambda_{cd}d$. Using the fact that $a+b+c+d=0$ and $\sum\lambda=1$, we can rewrite (\ref{ttt}) in the form
\begin{eqnarray}
&&-12\int \rmd^4p\rmd^4\lambda\delta(1-\lambda_{ab}-\lambda_{bc}-\lambda_{cd}-
\lambda_{da})\times\nonumber\\
&&(p\cdot p-m^2+Q+i\epsilon)^{-4}\times\nonumber\\
&&\mathrm{Tr}(\slashed{p}-\slashed{p}_{da}+m)\gamma^\alpha(\slashed{p}-\slashed{p}_{ab}+m)\gamma^\beta
(\slashed{p}-\slashed{p}_{bc}+m)\times\nonumber\\
&&\gamma^\gamma(\slashed{p}-\slashed{p}_{cd}+m)\gamma^\delta,\label{tt1}
\end{eqnarray}
where $Q$ is equal to
\begin{eqnarray}
&&(a\cdot a)\lambda_{da}\lambda_{ab}+(b\cdot b)\lambda_{ab}\lambda_{bc}\nonumber\\
&&+(c\cdot c)\lambda_{bc}\lambda_{cd}+
(d\cdot d)\lambda_{cd}\lambda_{da}\\
&&-(a+b)\cdot(c+d)\lambda_{bc}\lambda_{da}-(d+a)\cdot(b+c)\lambda_{ab}\lambda_{cd}\nonumber
\end{eqnarray}
and
\begin{eqnarray}
&&p_{da}=\lambda_{ab}a+\lambda_{bc}(a+b)-\lambda_{cd}d,\nonumber\\
&&p_{ab}=\lambda_{bc}b+\lambda_{cd}(b+c)-\lambda_{da}a,\nonumber\\
&&p_{bc}=\lambda_{cd}c+\lambda_{da}(c+d)-\lambda_{ab}b,\\
&&p_{cd}=\lambda_{da}d+\lambda_{ab}(d+a)-\lambda_{bc}c.\nonumber
\end{eqnarray}
In heads we need four factors of $abcd$ so, for their determination, we can drop $p$ and $m$ in the numerator. Then we can perform the trace in the numerator, leaving only heads. We can drop $p$ because it cannot appear in heads as from relativistic invariance we have $4p^\alpha p^\beta\to (p\cdot p)g^{\alpha\beta}$ and 
$24p^\alpha p^\beta p^\gamma p^\delta\to (p\cdot p)^2(g^{\alpha\beta}g^{\gamma\delta}+g^{\alpha\gamma}g^{\beta\delta}+g^{\alpha\delta}g^{\beta\gamma})$.
We get the head part of the trace in (\ref{tt1}) in the form $4\times$
\begin{eqnarray}
&&(p_{ab}^\alpha p_{bc}^\beta+p_{bc}^\alpha p_{ab}^\beta)(p_{cd}^\gamma p_{da}^\delta+p_{da}^\gamma p_{cd}^\delta)\nonumber\\
&&
+(p_{ab}^\alpha p_{cd}^\beta+p_{cd}^\alpha p_{ab}^\beta)(p_{bc}^\gamma p_{da}^\delta-p_{da}^\gamma p_{bc}^\delta)+\nonumber\\
&&(p_{ab}^\alpha p_{da}^\beta+p_{da}^\alpha p_{ab}^\beta)(p_{bc}^\gamma p_{cd}^\delta+p_{cd}^\gamma p_{bc}^\delta)\nonumber\\
&&+
(p_{bc}^\alpha p_{cd}^\beta-p_{cd}^\alpha p_{bc}^\beta)(p_{da}^\gamma p_{ab}^\delta-p_{ab}^\gamma p_{da}^\delta)\\
&&+(p_{da}^\alpha p_{bc}^\beta-p_{bc}^\alpha p_{da}^\beta)(p_{ab}^\gamma p_{cd}^\delta+p_{cd}^\gamma p_{ab}^\delta)\nonumber\\
&&+
(p_{cd}^\alpha p_{da}^\beta-p_{da}^\alpha p_{cd}^\beta)(p_{ab}^\gamma p_{bc}^\delta-p_{bc}^\gamma p_{ab}^\delta).\nonumber
\end{eqnarray}
Expanding the above expression we can find all heads. We shall only find heads of type 1 and 2. Type 1 is
\begin{widetext}
\begin{eqnarray}
&&32b^\alpha a^\beta d^\gamma c^\delta\lambda_{ab}\lambda_{cd}(1-\lambda_{ab})(1-\lambda_{cd}),32d^\alpha c^\beta b^\gamma a^\delta\lambda_{bc}\lambda_{da}(1-\lambda_{bc})(1-\lambda_{da}),\nonumber\\
&&\label{ty1}32c^\alpha d^\beta a^\gamma b^\delta
(\lambda_{ab}+\lambda_{bc})(\lambda_{bc}+\lambda_{cd})(\lambda_{cd}+\lambda_{da})(\lambda_{da}+\lambda_{ab}),
\end{eqnarray}
and type 2 is
\begin{eqnarray}
&&4(b^\alpha c^\beta d^\gamma a^\delta+d^\alpha a^\beta b^\gamma c^\delta)
(((1-\lambda_{da})(1-\lambda_{cd})+\lambda_{cd}\lambda_{da})((1-\lambda_{ab})(1-\lambda_{bc})+\lambda_{ab}\lambda_{bc})\nonumber\\
&&+((1-\lambda_{da})
\lambda_{cd}+(1-\lambda_{cd})\lambda_{da})((1-\lambda_{bc})\lambda_{ab}+(1-\lambda_{ab})\lambda_{bc}))-4(c^\alpha a^\beta d^\gamma b^\delta+b^\alpha d^\beta a^\gamma c^\delta)(((\lambda_{bc}+\lambda_{cd})(\lambda_{cd}+\lambda_{da})\nonumber\\
&&+(\lambda_{da}+\lambda_{ab})(\lambda_{ab}+\lambda_{bc}))(\lambda_{ab}\lambda_{cd}+(1-\lambda_{ab})(1-\lambda_{cd}))+
((\lambda_{cd}+\lambda_{da})(\lambda_{da}+\lambda_{ab})+(\lambda_{ab}+\lambda_{bc})(\lambda_{bc}+\lambda_{cd}))\nonumber\\
&&\times(\lambda_{ab}(1-\lambda_{cd})+\lambda_{cd}(1-\lambda_{ab}))),\label{ty2}\\
&&-4(d^\alpha c^\beta a^\gamma b^\delta+c^\alpha d^\beta b^\gamma a^\delta)
(((\lambda_{cd}+\lambda_{da})(\lambda_{da}+\lambda_{ab})+(\lambda_{ab}+\lambda_{bc})(\lambda_{bc}+\lambda_{cd}))(\lambda_{bc}\lambda_{da}+(1-\lambda_{bc})(1-\lambda_{da}))\nonumber\\
&&+
((\lambda_{bc}+\lambda_{cd})(\lambda_{cd}+\lambda_{da})+(\lambda_{da}+\lambda_{ab})(\lambda_{ab}+\lambda_{bc}))(\lambda_{bc}(1-\lambda_{da})+\lambda_{da}(1-\lambda_{bc}))),\nonumber
\end{eqnarray}
\end{widetext}
depicted in Fig. \ref{hour}. We also find that heads 5 appear in antisymmetric pairs, e.g., $(b^\alpha c^\beta a^\gamma-c^\alpha a^\beta b^\gamma)a^\delta$ and $(b^\alpha d^\beta a^\delta-d^\alpha a^\beta b^\delta)a^\gamma$, depicted in Fig. \ref{hour}.

To show that only tensors in (\ref{ten}) appear in $H$, note that (a) we can take away heads 1 and 2 by tensors 1 and 2 leaving the rest invariant,
(b) heads 3 and 4 can be taken away by tensors 3 and 4 leaving only heads 5 and 6, (c) heads 6, e.g., $b^\alpha a^\beta b^\gamma c^\delta$,
must get canceled when multiplied by $d_\delta$, which cannot be achieved either by terms with $g$ (they will contain $d$) or by other heads 6 or 5 (they give different terms). Therefore we only have to show that heads 5 alone must combine to the tensor 5. Let us focus first focus on a subgroup of these heads
$
(Ra^\delta+Sb^\delta+Tc^\delta)(b^\alpha c^\beta a^\gamma-c^\alpha a^\beta b^\gamma)
$. 
Multiplied by $d$ cycles $b^\alpha c^\beta a^\gamma-c^\alpha a^\beta b^\gamma$ must get canceled from gauge invariance, which cannot be done by terms with $g$ ($d$ 
will appear) or other groups (we cannot get the same cycles). Hence $R(a\cdot d)+S(b\cdot d)+T(c\cdot d)=0$ and we can rewrite the group
multiplied by $(c\cdot d)$ in the form 
$(R((c \cdot d)a^\delta-(a\cdot d)c^\delta)+S((c\cdot d)b^\delta-(b\cdot d)c^\delta))(b^\alpha c^\beta a^\gamma-c^\alpha a^\beta b^\gamma)$. which corresponds to a combination of tensors 5.

Making the Wick rotation $p^0\to \rmi p^0$, we can also integrate (\ref{tt1}) over $p$,
\begin{eqnarray}
&&\int \frac{\rmd^4p}{(p\cdot p-m^2+Q+\rmi\epsilon)^{4}}=\int_0^\infty \frac{2\pi^2\rmi P^3\rmd P}{(P^2+m^2-Q-\rmi\epsilon)^4}
\nonumber\\
&&=\int_0^\infty\frac{\pi^2\rmi u\rmd u}{(u+m^2-Q-\rmi\epsilon)^4}=\frac{\rmi\pi^2}{6(m^2-Q-\rmi\epsilon)^2}.\label{tt2}
\end{eqnarray}

At small values of $abcd$ we can neglect $Q$ in the denominator of (\ref{tt2}).
It remains to integrate (\ref{ty1}) and (\ref{ty2}) (the other tensors appear only at large values as they must contain additional Fourier variables $abcd$ from $Q$) over $\lambda$, and the final result is
\begin{eqnarray}
&&H=-\frac{8\rmi\pi^2}{9m^4}((\cdot A\cdot B\cdot)(\cdot C\cdot D\cdot)+\nonumber\\
&&(\cdot A\cdot C\cdot)(\cdot B\cdot D\cdot)+(\cdot A\cdot D\cdot)(\cdot C\cdot B\cdot))\nonumber\\
&&+\frac{28\rmi\pi^2}{45m^4}((\cdot A\cdot B\cdot C \cdot D\cdot)+\nonumber\\
&&(\cdot A\cdot C\cdot B \cdot D\cdot)+(\cdot A\cdot B\cdot D \cdot C\cdot)).
\end{eqnarray}


\begin{thebibliography}{99}
\bibitem{bell}
J.S. Bell, Physics (Long Island City, N.Y.) \textbf{1}, 195 (1964);
A. Shimony, plato.stanford.edu/entries/bell-theorem/.
 \bibitem{chsh}
J.F. Clauser, M.A. Horne, A. Shimony and R.A. Holt,
Phys. Rev. Lett. \textbf{23}, 880(1969) .
\bibitem{eber}
P.H. Eberhard, Phys. Rev. A \textbf{47}, R747 (1993).
\bibitem{hanson}
B. Hensen \emph{et al.}, Nature (London)\textbf{526}, 682 (2015)
%\bibitem{hensenraw}
%dx.doi.org/10.4121/uuid:6e19e9b2-4a2d-40b5-8dd3-a660bf3c0a31
\bibitem{zein}
M. Giustina \emph{et al.}, Phys. Rev. Lett. \textbf{115}, 250401 (2015).
\bibitem{saew}
L.K. Shalm \emph{et al.}, Phys. Rev. Lett. \textbf{115}, 250402 (2015).
\bibitem{wight}
R. F. Streater and A. S. Wightman \emph{PCT, Spin and Statistics, and All that}
 (Benjamin, New York, 1964).
\bibitem{peskin}
N. Bogoliubov, A.A. Logunov, A.I. Oksak, and I.T. Todorov, \textit{General Principles of Quantum Field Theory}, (Kluwer Academic Publishers, Dordrecht, 1990); L.H. Ryder, \emph{Quantum Field Theory} (Cambridge University Press, Cambridge, 1985);
S. Weinberg, \textit{The Quantum Theory of Fields} (Cambridge University Press, Cambridge, 1995);
M. Peskin, D. Schroeder, \textit{An Introduction to Quantum Field Theory} (Perseus Books, Reading, 1995);
J.D. Bjorken and S.D. Drell, \textit{Relativistic Quantum Mechanics} (McGraw-Hill, New York, 1998).
\bibitem{ab13}
A. Bednorz, Eur. Phys. J. C \textbf{73}, 2654 (2013) .
\bibitem{povm} H.M. Wiseman and  G.J. Milburn,  \textit{Quantum
  Measurement and Control} (Cambridge University Press, Cambridge, England, 2009).
\bibitem{kraus} 
K. Kraus, \textit{States, Effects and Operations}
(Springer, Berlin, 1983)  
\bibitem{grw}
G.C. Ghirardi, A. Rimini and T. Weber, Phys. Rev. D \textbf{34}, 470 (1986).
 \bibitem{csl}
P. Pearle, Phys. Rev. A  \textbf{39}, 2277 (1989);
G.C. Ghirardi, P. Pearle, and A. Rimini, Phys. Rev. A \textbf{42}, 78 (1990);
G.C. Ghirardi, R. Grassi, and P. Pearle, Found. Phys. \textbf{20}, 1271 (1990).
\bibitem{weak} Y. Aharonov, D.Z. Albert and L. Vaidman, Phys. Rev. Lett. \textbf{60}, 1351 (1988). 
\bibitem{bb10} A. Bednorz and W. Belzig, Phys. Rev. Lett. \textbf{105}, 106803 (2010).
\bibitem{abn} A. Bednorz, W. Belzig and A. Nitzan, New J. Phys. \textbf{14}, 013009 (2012).
\bibitem{bfb}
A. Bednorz, K. Franke, and W. Belzig, New J. Phys. \textbf{15}, 023043 (2013).
 \bibitem{wigner}
E.P. Wigner, Phys. Rev. \textbf{40}, 749 (1932);
M. Hillery \emph{et al.}, Phys. Rep. {\bf 106}, 121 (1984).
\bibitem{ab15}
A. Bednorz, New J. Phys. \textbf{17}, 093006 (2015).
\bibitem{clerk}
A.A. Clerk, M.H. Devoret, S.M. Girvin, F. Marquardt, and R.J. Schoelkopf, Rev. Mod. Phys. \textbf{82}, 1155 (2010).
\bibitem{bbrb}
A. Bednorz, C. Bruder, B. Reulet and W. Belzig, Phys. Rev. Lett. \textbf{110}, 250404 (2013).
\bibitem{zwanzig} R. Zwanzig, Physica \textbf{30}, 1109 (1964).
\bibitem{ctp0}
J. Schwinger, J. Math. Phys. {\bf 2}, 407 (1961);
L.P. Kadanoff and G. Baym, \textit{Quantum Statistical Mechanics}, (W.A. Benjamin, New York, 1962);
L. Keldysh, Sov. Phys. JETP \textbf{20}, 1018 (1965).
\bibitem{ctp}
K. Chou, Z. Su, B. Hao and L. Yu, Phys. Rep. \textbf{118}, 1 (1985);
N.P. Landsman and C.G. van Weert, Phys. Rep. \textbf{145}, 141 (1987).
\bibitem{bbb}
A. Bednorz, W. Bednorz, and W. Belzig, Phys. Rev. A \textbf{89}, 022125 (2014).
\bibitem{fdt} 
H.B. Callen and T.A. Welton, Phys. Rev. \textbf{83}, 34 (1951)
\bibitem{g44}
R. Karplus and M. Neuman, Phys. Rev. \textbf{80}, 380 (1950).
\bibitem{kost}
A. Kostelecky and N. Russell, Rev. Mod. Phys. \textbf{83}, 11 (2011).
\bibitem{wise}
H. Wiseman, Nature (London) \textbf{510}, 467 (2014).
\end{thebibliography}
\end{document}